\documentclass[referee]{aa}
\usepackage{graphicx}

\begin{document}

\newcommand{\etal}{ {\it et al.}}
\newcommand{\porb}{ P_{orb} } 
\newcommand{\Po}{$ P_{orb} \su$}
\newcommand{\pdot}{$ \dot{P}_{orb} \,$}
\newcommand{\pot}{$ \dot{P}_{orb} / P_{orb} \su $}
\newcommand{\mm}{$ \dot{m}$ }
\newcommand{\mdot}{$ |\dot{m}|_{rad}$ }
\newcommand{\myr}{ \su M_{\odot} \su \rm yr^{-1}}
\newcommand{\msol}{\, M_{\odot}}
\newcommand{\ppp}{ \dot{P}_{-20} }
\newcommand{\cms}{ \rm \, cm^{-2} \, s^{-1} }
\newcommand{\pdott}{ \left( \frac{ \dot{P}/\dot{P}_o}{P_{1.6}^{3}} \right)}

\def\p{\phantom{1}}
\def\pmu{\mox{$^{-1}$}}
\def\ApJ{{\it Ap.\,J.\/}}
\def\ApJL{{\it Ap.\,J.\ (Letters)\/}}
\def\ApJS{{\it Astrophys.\,J.\ Supp.\/}}
\def\AJ{{\it Astron.\,J.\/}}
\def\AA{{\it Astr.\,Astrophys.\/}}
\def\AAL{{\it Astr.\,Astrophys.\ Letters\/}}
\def\AAS{{\it Astr.\,Astrophys.\ Suppl.\,Ser.\/}}
\def\MN{{\it Mon.\,Not.\,R.\,Astr.\,Soc.\/}}
\def\Na{{\it Nature \/}}
\def\SAIt{{\it Mem.\,Soc.\,Astron.\,It.\/}}
\def\kms{km^s$^{-1}$}
\def\sbu{mag^arcsec${{-2}$}}
\def\e{\mbox{e}}
\def\dex{\mbox{dex}}
\def\L{\mbox{${\cal L}$}}
\def\gte{\lower 0.5ex\hbox{${}\buildrel>\over\sim{}$}}
\def\lte{\lower 0.5ex\hbox{${}\buildrel<\over\sim{}$}}
\def\loe{\lower 0.6ex\hbox{${}\stackrel{<}{\sim}{}$}}
\def\goe{\lower 0.6ex\hbox{${}\stackrel{>}{\sim}{}$}}

\title{GRB 990704: the most X-ray rich BeppoSAX gamma-ray burst}

\author{
M. Feroci\inst{1}, 
L.A. Antonelli\inst{2},
P. Soffitta\inst{1},
J.J.M. in 't Zand\inst{3},
L. Amati\inst{4},
E. Costa\inst{1},
L. Piro\inst{1},
F. Frontera\inst{4,5},
E. Pian\inst{4,6},
J. Heise\inst{3},
L. Nicastro\inst{7}
}

\offprints{feroci@ias.rm.cnr.it}

\institute{
{Istituto Astrofisica Spaziale, C.N.R.,
  Area di Tor Vergata, Via Fosso del Cavaliere 100,
  00133 Roma, Italy}
\and
{Osservatorio Astronomico di Roma, Via Frascati 33, 00040 Monteporzio Catone, Italy}
\and
{Space Research Organization in the Netherlands, Sorbonnelaan 2, 3584, CA Utrecht, The Netherlands}
\and
{Istituto Tecnologie e Studio Radiazioni Extraterrestri,
CNR, Via Gobetti 101, 40129 Bologna, Italy}
\and
{Dipartimento di Fisica, Universit\`a di Ferrara, Via Paradiso
 12, 44100 Ferrara, Italy}
\and
{Osservatorio Astronomico di Trieste, Via G.B. Tiepolo 11, I-34131, Trieste,
Italy}
\and
{Istituto di Fisica Cosmica ed Applicazioni dell'Informatica,
CNR, Via Ugo La Malfa 153, 90138 Palermo, Italy}
}

\date{Received 30 March 2001; accepted 27 August 2001}

\abstract{
We present the X- and $\gamma$-ray detection of GRB 990704 
and the discovery and study of its X-ray afterglow, 1SAX J1219.5-0350. 
Two pointed BeppoSAX observations with the narrow field instruments
were performed on this source, separated 
in time by one week. The decay of the X-ray flux within the first 
observation appears unusually slow, being best-fit by a power law 
with negative index 0.83$\pm$0.16.
Such a slow decay is consistent with the non-detection in our second
observation, but its
back-extrapolation to the time of the GRB largely underestimates
the detected GRB X-ray prompt emission. 
In addition, the GRB prompt event shows, among the BeppoSAX-WFC detected
sample, unprecedentedly high ratios of X- and gamma-ray peak fluxes 
(F$_{2-10~keV}$/F$_{40-700~keV}$$\sim$0.6, and
F$_{2-26~keV}$/F$_{40-700~keV}$$\sim$1.6)
and fluences (S$_{2-10~keV}$/S$_{40-700~keV}$$\sim$1.5 and
S$_{2-26~keV}$/S$_{40-700~keV}$$\sim$2.8), making it, among the
BeppoSAX arcminute-localized GRBs, the closest to the recently discovered 
class of Fast X-ray Transients.
}

\titlerunning{GRB 990704: the most X-ray rich BeppoSAX GRB}
\authorrunning{M. Feroci et al.}

\maketitle

\section{Introduction}

Despite the huge step forward allowed by the BeppoSAX discovery
of the X-ray afterglows (\cite{costa97}) of gamma-ray bursts (GRBs)
and by the related discoveries of optical (\cite{vanparadijs97}) and
radio afterglows (\cite{frail97}), the ultimate explanation of GRBs 
remains unknown.  
Several sources associated with GRBs have been demonstrated to be at 
cosmological distances (e.g., see \cite{costa99} and \cite{lamb00}
for reviews),
but the current sample only includes long duration GRBs 
leaving open the possibility of the existence of different populations
of GRB sources (e.g., \cite{tavani98} and references therein). 

The observations of new GRB afterglows continue 
to show new properties that contribute to making the interpretative 
scenario complex and incomplete. 
An interesting example is the so-called GRB-supernova connection.
After the discovery by BeppoSAX of a GRB
possibly associated with a type Ic supernova
(GRB 980425, \cite{pian99,galama98}),
the possibility that some GRBs might be associated with supernovae
has been proposed and supported by late-time observations
of GRBs like 980326, 970228, 990712 and 970508 
(\cite{bloom99,reichart99,hjorth99,sokolov01}).
The case for GRB 990712 is now debated (\cite{sahu00,hjorth00,bjornsson01}). 

A further interesting result recently provided by BeppoSAX is 
the identification of a potential new sub-class of GRBs: 
the Fast X-ray Transients (FXT, \cite{heise01}).
They are flashes of X-rays, so far not recurrent, sometimes
accompanied by weak gamma-ray emission, that
show no properties common to any known class of X-ray sources. 
The unique property of these events
is their large X--ray content, comparable to or dominant over their emission
at gamma-rays. 
Their occurrence rate is approximately one third of the GRBs;
they could be 
X-ray counterparts of a - so far unexplored - 
new class of very soft GRBs (\cite{kippen01,heise01}).
Attempts to find afterglows have failed so far.
Fascinating proposals have been suggested to interpret
them, going from `dirty fireball' GRBs (e.g., Dermer 1999) 
to highly redshifted classical GRBs.

In this paper we present the detection, localization and study of  
GRB 990704 and its X--ray afterglow. This event, the 17th GRB 
promptly localized by the BeppoSAX Wide Field Cameras
and soon after observed with the BeppoSAX Narrow Field Instruments,
is peculiar in many respects, and may be related to the FXT class.

\section{GRBM and WFC Observations}

The gamma-ray burst GRB 990704 triggered the Gamma Ray Burst Monitor
(GRBM, \cite{frontera97,feroci97,costa98}) onboard BeppoSAX on 1999, 
July 4th 17:30:20.221 UT (63020.221 seconds of day, SOD)
and was at the same time detected and imaged by
unit 1 of the Wide Field Cameras (WFC, \cite{jager97}).
WFC data revealed the GRB 
with a signal-to-noise ratio of 18 at the position
in the sky (J2000) R.A.=12$^{h}$19$^{m}$30$^{s}$ and 
Decl.=-3$^{\circ}$48'.2, with an error radius (99\% confidence level)
of 7' (\cite{heise99}).
The relatively large size of the WFC error region is primarily
due to an unfavorable satellite attitude.
An independent position of the same event was obtained
through the analysis of the difference in the arrival times of the event
at the BeppoSAX GRBM and the Ulysses GRB detector. This resulted
in a 2'.8 wide (3-$\sigma$) annulus in the sky intersecting the 
WFC error circle, which reduced the error box to   
$\sim$75 arcmin$^{2}$ (see Figure~\ref{due_grb}) (\cite{hurley99}).

\begin{figure}
\centerline{\includegraphics[width=8.5cm]{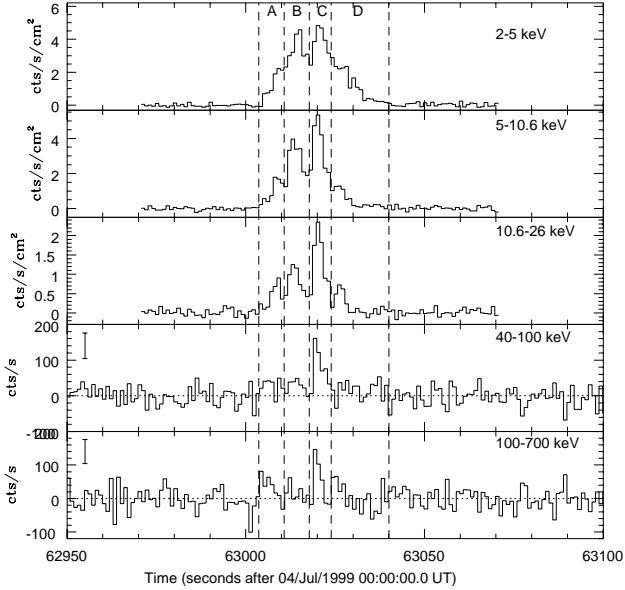}}
\caption[]{
WFC (top 3 panels) and GRBM (bottom 2 panels, on the top left the typical
$\pm$1$\sigma$ uncertainty is given) light curves of GRB 990704. 
Vertical dashed lines identify the intervals for the time-resolved
spectral analysis (see text). 
 }
\label{lc_wfc_grbm}
\end{figure}

The light curve of the event as derived from the GRBM and WFC experiments
is shown in different energy ranges in Figure~\ref{lc_wfc_grbm}. 
The GRBM light curve is given in two separate energy ranges 
(40--100~keV and 100--700~keV). 
They are derived from the available GRBM data:
1-s count-rates in the two partially overlapping ranges 40--700~keV 
and $>$100~keV.
Since the GRBM effective area above 700~keV is very small, 
we can usually assume that the contribution to the count rate from photons 
with energy above 700~keV is negligible, and we can obtain light curves 
in the two adjacent ranges 40--100~keV and 100--700~keV
(see \cite{amati99} for an extensive discussion on this subject).  
This procedure has been verified in the specific case of GRB 990704
with the time-averaged spectral data available from the GRBM (e.g.,
\cite{feroci97}). 
The 240-channel energy spectra, in the full energy range from 40 to 700 keV,
integrated over fixed 128-s time intervals, were used to derive
the time-averaged spectral shape extrapolation beyond 700 keV, 
verifying
that the number of counts expected above 700 keV is negligible with respect 
to those detected in 40-100 keV. In the case of GRB 990704 we expect the 
counts above 700 keV to be less than 0.3\% of those detected in the 
40-100 keV range.

\begin{table}
\label{tab1}
\begin{center}
\caption{Peak fluxes (1-s) and fluences of GRB 990704 in different energy ranges.}
\begin{tabular}{|l|l|l|l|}
\hline
             & 2-10 keV      &  2-26 keV        & 40-700 keV  \\ \hline
Peak Flux, F & $1.0\pm0.1$   & $2.80\pm0.28$    & $1.8\pm0.2$ \\ 
(10$^{-7}$ erg cm$^{-2}$ s$^{-1}$) &  &      &             \\ \hline
Fluence, S   & $1.52\pm0.08$ &  $2.84\pm0.14$   & $1.0\pm0.1$ \\ 
(10$^{-6}$ erg cm$^{-2}$ )   &  &     &             \\ \hline
\end{tabular}
\end{center}
\end{table}

\begin{figure}
\includegraphics[width=6.5cm]{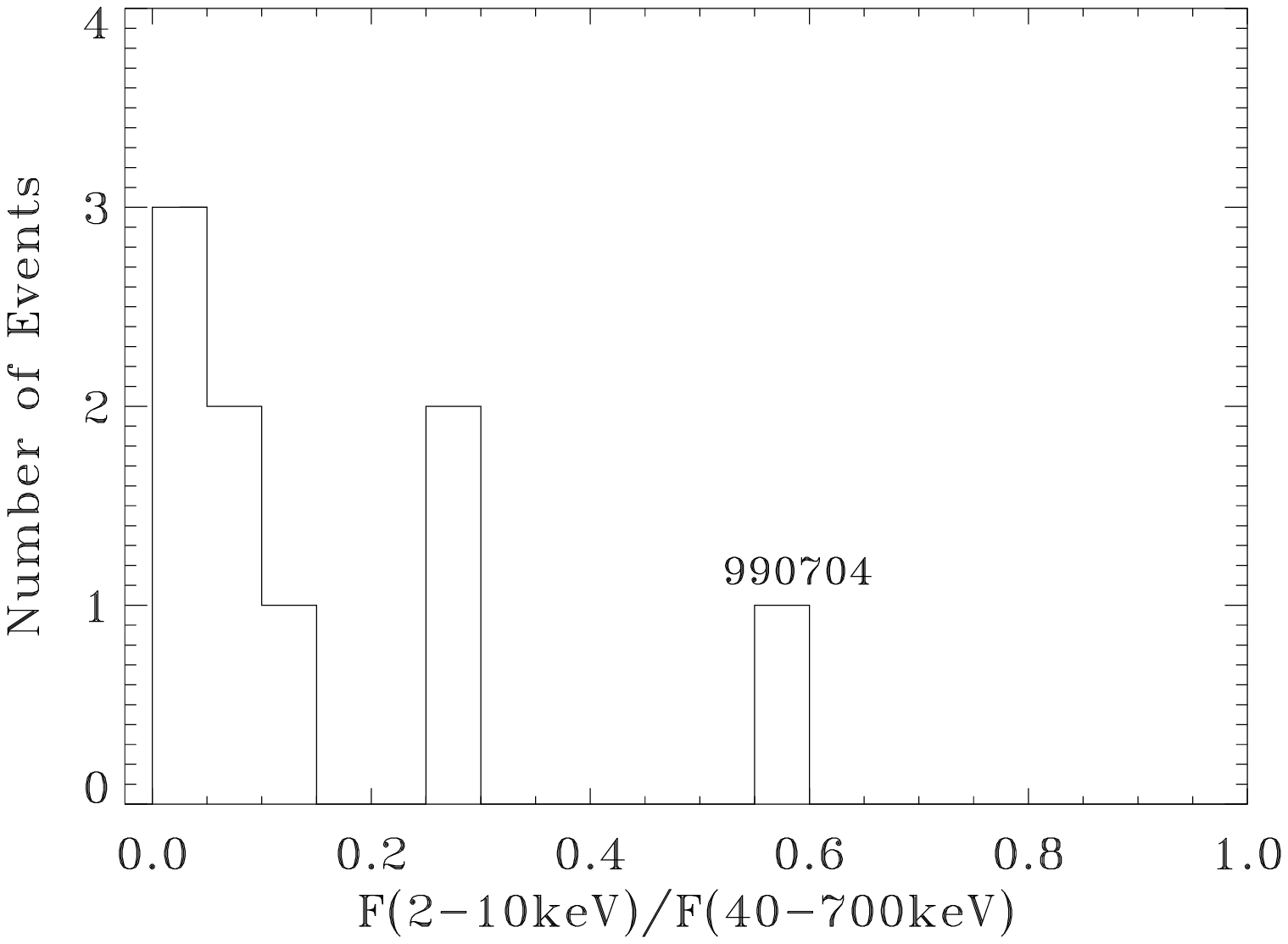}
\includegraphics[width=6.5cm]{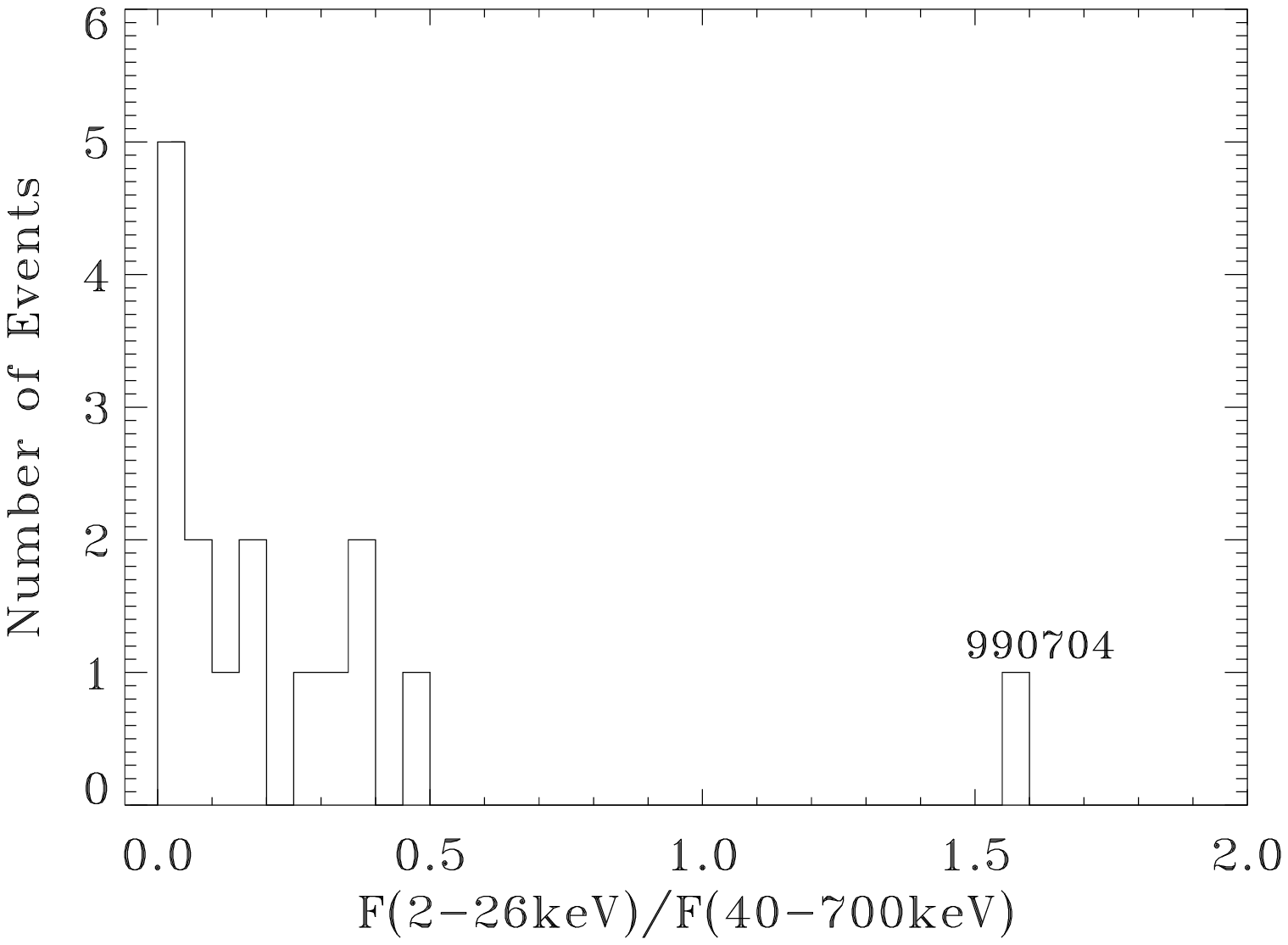}
\caption{
Histogram of the X-to-gamma peak flux ratios, for the two X-ray energy ranges 2-10 
and 2-26 keV, of the same events used to derive Table 2.
}
\label{ratios_p}
\end{figure}

\begin{figure}
\includegraphics[width=6.5cm]{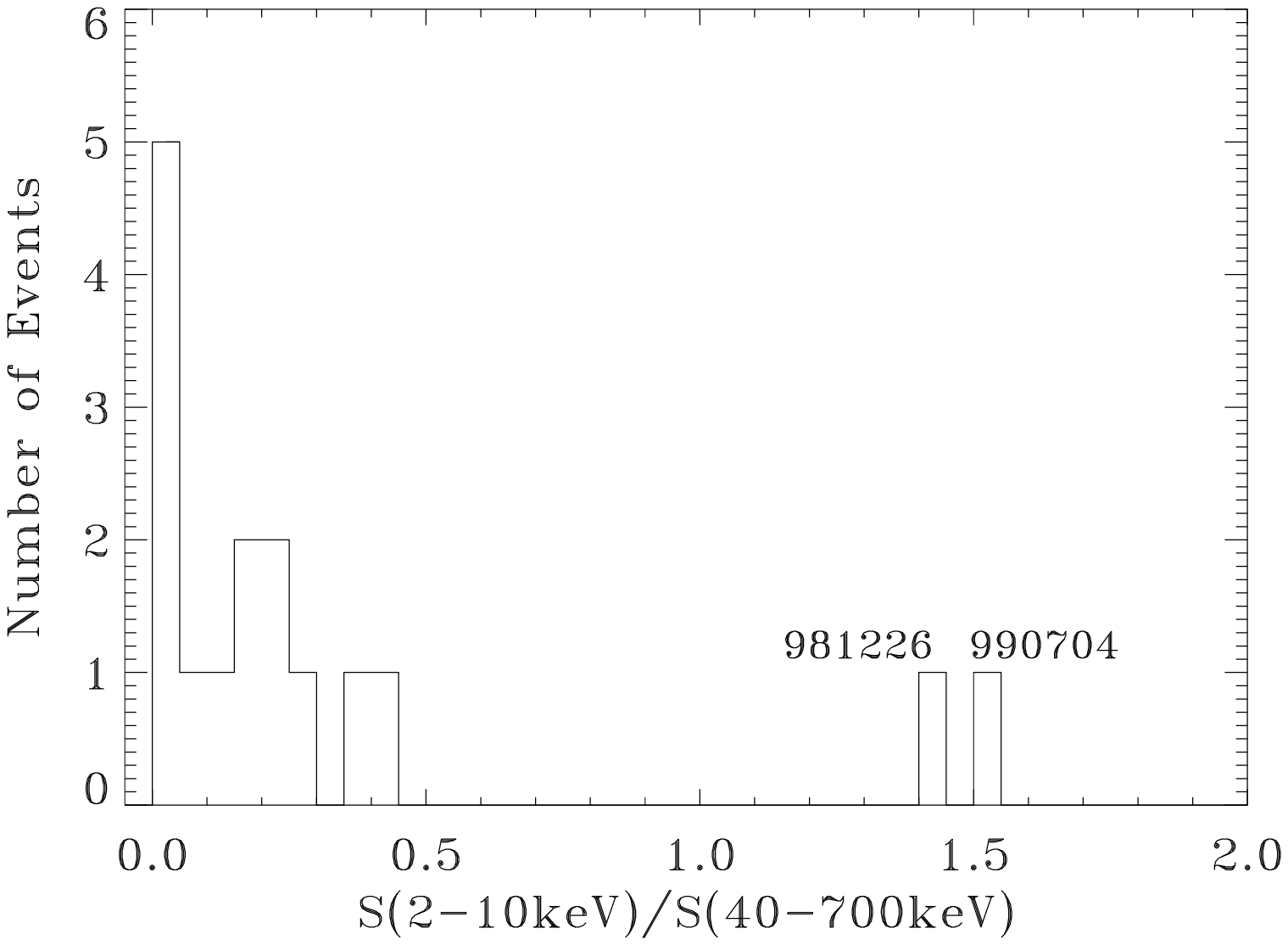}
\includegraphics[width=6.5cm]{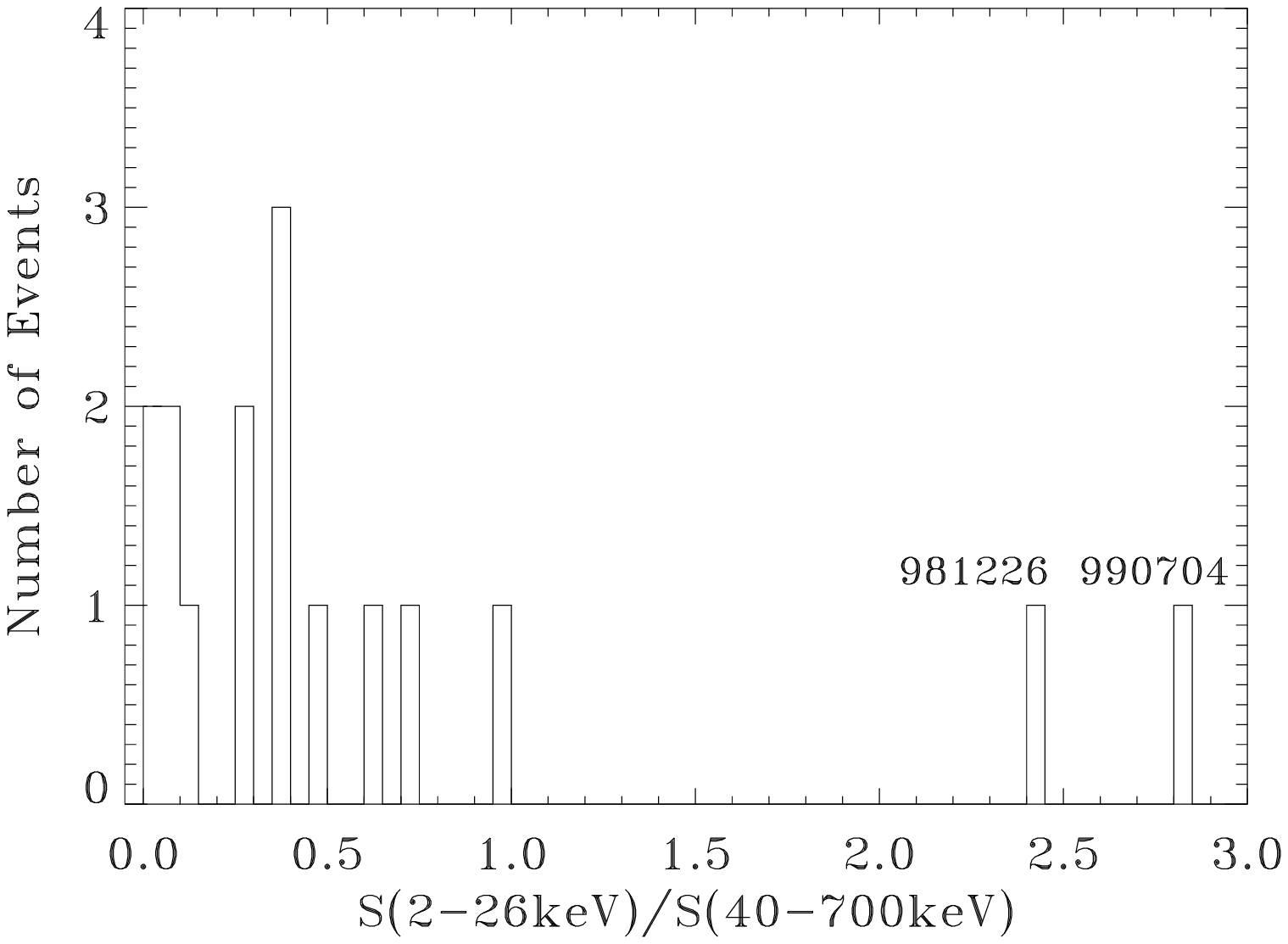}
\caption{
X-to-gamma fluence ratios, for the two X-ray energy ranges 2-10
and 2-26 keV, of the same events used to derive Table 2.
}
\label{ratios_s}
\end{figure}

The peak fluxes and the fluences for GRB 990704 in different energy ranges
are given in Table 1. They can be used to compute the ratio of peak fluxes,
namely 
F$_{X}$/F$_{\gamma}$ = 0.56$\pm$0.09 using the 2-10~keV X-ray flux
and 1.56$\pm$0.23 using the 2-26~keV X-ray flux: the largest
values found so far in the BeppoSAX sample (Figure~\ref{ratios_p}).
The duration of the event is perhaps slightly energy dependent, ranging
from $\sim$37~s in 2--5~keV to less than $\sim$30~s above 100~keV.
The ratios of fluences in X-rays and gamma-rays are again 
larger than any in the BeppoSAX sample (see Figure~\ref{ratios_s}):
S$_{2-10~keV}$/S$_{40-700~keV}$ = 1.52$\pm$0.15 and
S$_{2-26~keV}$/S$_{40-700~keV}$ = 2.84$\pm$0.27.
As a comparison, if we consider the sample of the GRBs
localized by BeppoSAX between July 1996 and December 1999, we 
obtain the average values and scatters provided in Table 2. 
The basic data for the computation were available only for
subsets of the total number of events, specified in parentheses.
A systematic analysis of the complete sample will be 
reported elsewhere (Frontera et al., in preparation). 
An examination of the quick-look data  
(that is, the raw data promptly analysed at the 
BeppoSAX Science Operation Center)
for the events not included in our analysis, indicates that
their exclusion should not significantly affect our results.
Table 2 makes it very evident that 
GRB 990704 and 981226 are both responsible for a large increase of both the
average and the scatter of the fluence ratios, whereas
990704 alone causes a significant increase in the average and scatter
of the peak flux ratios.

\begin{table}
\label{tab2}
\begin{center}
\caption{
The average, $<x>$, and the standard deviation,
$s$, of the X to gamma content of the sample of GRBs localized by BeppoSAX
between 1996 and 1999. The total number of events is 28. For each of the
estimates, the number of events for which the data were available is 
given in parentheses (see also text). 
}
\begin{tabular}{|l|l|l|l|l|l|}
\hline
          & F$_{2-10~keV}$/ & F$_{2-26~keV}$/ & S$_{2-10~keV}$/ & S$_{2-26~keV}$/ & Notes \\ 
          & F$_{40-700~keV}$ & F$_{40-700~keV}$ & S$_{40-700~keV}$ & S$_{40-700~keV}$ &  \\ \hline
$<x>$    & 0.15(9) & 0.29(16) & 0.31(17) & 0.63(16) & \\  \hline
$s$      & 0.18(9) & 0.42(16) & 0.45(17) & 0.83(16) & \\  \hline
$<x>$    & 0.10(8) & 0.20(15) & 0.23(16) & 0.47(15) & excluding 990704 \\ \hline
$s$      & 0.10(8) & 0.26(15) & 0.33(16) & 0.60(15) & excluding 990704 \\ \hline
$<x>$    & 0.08(7) & 0.18(14) & 0.15(15) & 0.34(14) & excluding 990704 \\  
         &         &          &          &          & and 981226       \\ \hline
$s$      & 0.08(7) & 0.26(14) & 0.13(15) & 0.28(14) & excluding 990704 \\ 
         &         &          &          &          & and 981226       \\ \hline

\end{tabular}
\end{center}
\end{table}

The WFC did not detect X--ray emission from the 
location of GRB 990704 after the end of the GRB prompt emission. 
The 3-$\sigma$ upper limit in the 2--10~keV range in the time interval 
63040-63076~s UT (that is, the 36 s after the end of the GRB)
is 1.6$\times10^{-9}$ erg cm$^{-2}$ s$^{-1}$.

\begin{figure}
\centerline{\includegraphics[width=8.5cm]{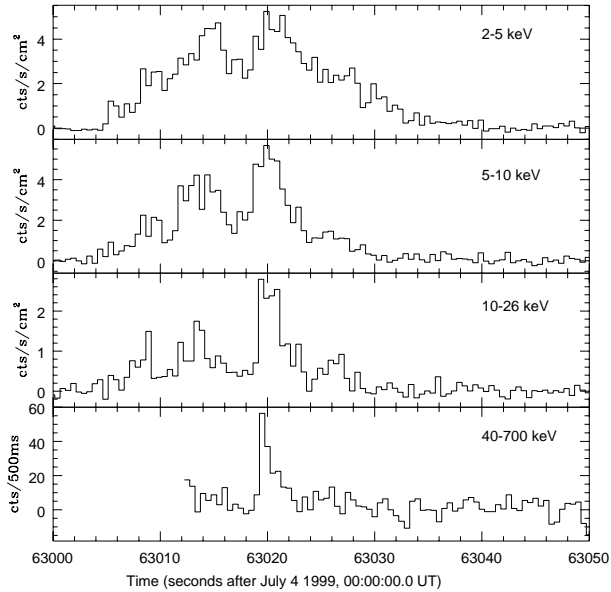}}
\caption{
Light curve of GRB 990704 with 500~ms time resolution, providing a 
clear view of the energy dependence of the width of the pulse at 
63020 seconds UT. For the GRBM data, only the integrated 40-700~keV 
energy range is available with this time resolution.
}
\label{hrlc}
\end{figure}

\begin{figure}
\centerline{\includegraphics[width=8.5cm]{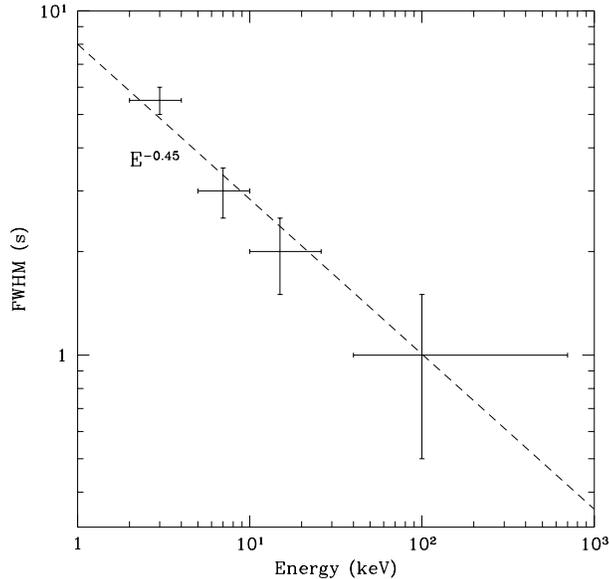}}
\caption{
Full Width at Half Maximum of the main peak of GRB990704 (see temporal
slice C in Fig.~\ref{lc_wfc_grbm}), as a function
of energy. The dashed line shows a FWHM $\propto$ E$^{-0.45}$ relation.
Horizontal
error bars indicate the energy band. Vertical error bars represent
500~ms (see text and Fig.~\ref{hrlc}), 
they must be taken only as an indication 
of the uncertainty on the duration estimate.
}
\label{width}
\end{figure}

Only one of the four X--ray peaks has
a significant gamma-ray counterpart, the one included in interval
C in Figure~\ref{lc_wfc_grbm}; it is also the most intense in the X-ray range. 
The shape of the gamma-ray pulse exhibits a profile consistent with
a FRED (Fast Rise Exponential Decay) profile. 
We checked the energy dependence of the width of this pulse.
For this purpose, we used the high time resolution data available
from both the WFC and the GRBM, binned at 0.5~s (Figure~\ref{hrlc}). 
We could derive a value of the duration (in terms of 
full width at half maximum, FWHM) of the main peak
only in four energy ranges because the high time resolution data from 
the GRBM
are only available in the integrated energy range 40-700 keV, and 
they start 8~s before the trigger time. 
The estimated FWHM values are shown in Figure~\ref{width}. We found
that the width uncertainties were comparable to the time resolution.
Although the multi-peaked structure observed at X-rays might 
significantly affect our estimates, 
the derived values appear in agreement with the $E^{-0.45}$ 
($E$ photon energy)
dependence found by Fenimore et al. (1995) for the 
classical GRBs (e.g., \cite{piro98} and references therein).

\subsection{Spectral Analysis}

We analyzed the energy spectrum of the GRB event using the
WFC and GRBM data. To perform a time-resolved spectral
analysis, for the GRBM we used the two-channel spectra that can be derived
from the two energy ranges of the 1-s ratemeters (\cite{amati99}). 
The two available channels cover the energies 40-100 and 100-700~keV.
We selected four time intervals (A, B, C, D; Figure~\ref{lc_wfc_grbm})
during the event, encompassing the
four distinct peaks visible in the energy range 11-26~keV in the
WFC data. The GRBM high energy (100-700~keV)
channel does not provide a significant detection in any of the four
selected intervals, and this datum was therefore excluded 
during the spectral fitting procedure.

We first attempted to model the spectrum in the four intervals with 
a simple (unabsorbed) power law. This solution is acceptable only for 
the interval D (see Table 3), for which intrinsic 
photoelectric absorption
is not needed. For the other intervals different models were needed.
In particular, the average spectrum of interval A can be satisfactorily 
fit either by adding photoelectric absorption by cold matter 
or changing the spectral model to a broken power law.  
In intervals B and C an absorbed power law is not acceptable, 
whereas the broken power law is satisfactory.

\begin{table}
\label{tab3}
\begin{center}
\caption{Time-resolved spectral analysis of GRB 990704. 
The spectral model is either a simple power law (photon index $\alpha$),
an absorbed power law, or a broken power law (low and high energy
photon indices $\alpha$ and $\beta$, break energy  E$_{break}$).
Errors are given at 90\% confidence level. Asterisks (*) indicate 
parameters for which an error could not be derived because of a large
$\chi^{2}$.}
\begin{tabular}{|l|l|l|l|l|l|}
\hline
          & $\alpha$               & E$_{break}$ (keV)     & $\beta$                & N$_{H}$$\times$10$^{22}$ cm$^{-2}$ & $\chi^{2}_{\nu}$(dof) \\ \hline
A         & $2.12^{+0.16}_{-0.14}$           &     -                 &    -                   & 4.85$^{+4.25}_{-3.14}$       & 0.93(13) \\ 
          & 1.31$^{+0.34}_{-0.98}$ & 11.4$^{+50.2}_{-7.1}$ & 2.35$^{+4.14}_{-0.32}$ &   -                          & 0.80(10) \\  \hline
B         & 2.55$^*$         & - & - & 8.7$^*$ & 2.16(12) \\  
          & 0.92$^{+0.30}_{-0.34}$ &  7.4$^{+ 2.3}_{-1.1}$ & 2.70$^{+0.21}_{-0.16}$ &   -                          & 1.04(11) \\  \hline 
C        & 2.14$^{+0.08}_{-0.07}$ &  -  &  -  & 4.23$^{+1.86}_{-1.61}$  & 1.43(12) \\
         & 1.29$^{+0.22}_{-0.27}$ &  9.3$^{+41.9}_{-2.3}$ & 2.29$^{+2.75}_{-0.13}$ &   -                          & 0.87(11) \\  \hline
D        & 2.14$^{+0.14}_{-0.11}$ & - & - & $<$2.55 & 1.15(12) \\
         & 2.11$^{+0.10}_{-0.09}$ & - & - &   -     & 1.09(13)  \\  \hline 
\end{tabular}
\end{center}
\end{table}

\section{The X-ray afterglow}

\subsection{Observations}

Soon after the localization of GRB 990704 by the WFC a
Target of Opportunity observation with the BeppoSAX Narrow Field Instruments
(NFI, \cite{boella97}) was initiated.
The NFI observation began on July 5, 01:48 UT, approximately
eight hours after the GRB, and was initially
scheduled to last for 100~ks of net observing time. However, following
the GRBM/WFC detection and localization of GRB 990705 
(e.g., \cite{amati00}) 
a new Target of Opportunity was declared on the latter GRB and the 
NFI observation of the field of GRB 990704 was stopped 
on July 5, 23:45 UT for a total net exposure time of 37 ks
for the MECS and 13.4 ks for the LECS. 
 
\begin{figure*}
\centering
\includegraphics[angle=270,width=13cm]{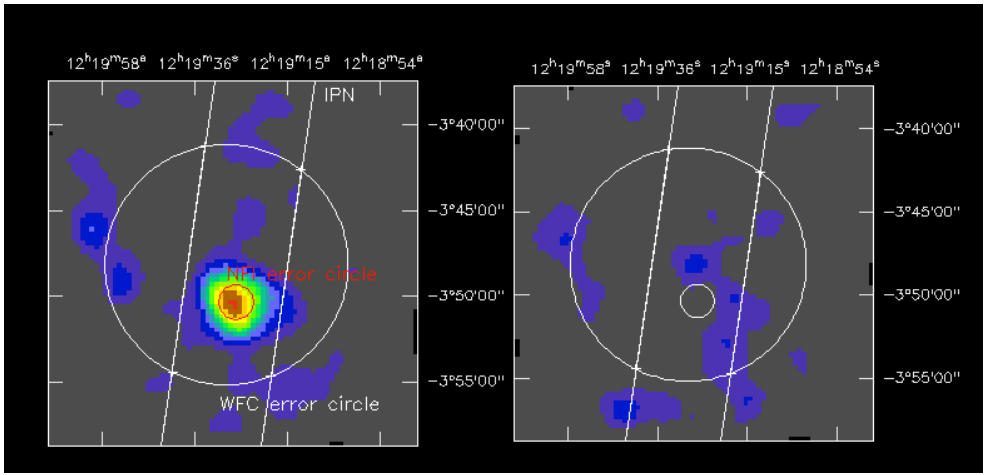}
\caption[]{
MECS image of the field of GRB 990704 8 hours after the burst
(left) and one week later (right). The fading X-ray source is
1SAX J1219.5-0350. The WFC, IPN and NFI error boxes are indicated. 
 }
\label{due_grb}
\end{figure*}

A previously unknown X-ray source, 1SAX J1219.5-0350, 
was detected almost at the center of the MECS and LECS detectors,
at the sky position (J2000) R.A.=12$^{h}$19$^{m}$27$^{s}$.3 and
Decl.=-3$^{\circ}$50'~22" (90\% confidence level error radius 1'),
consistent with the GRB 990704 combined error box
(Figure~\ref{due_grb}, \cite{feroci99b}; \cite{hurley99}).
The source flux, averaged over the entire NFI observation, was
(5.64$\pm$0.45) x 10$^{-3}$ counts s$^{-1}$ in the MECS (1.6--10~keV)
and (3.2$\pm$0.8) x 10$^{-3}$ counts s$^{-1}$ in the LECS (0.1--2~keV)
(uncertainties on the fluxes are 1-$\sigma$).

Given that the X-ray flux from 1SAX J1219.5-0350 appeared 
to decrease with time during the observation, a new NFI pointing
was scheduled at its position in order to verify its association with
GRB 990704. This second NFI observation started on July 11, 19:10 UT
and ended on July 12, 20:48 UT, for a net MECS exposure time of 42.5 ks.
We averaged the flux over both the entire second pointing and also
over its first 17~ks, and found no detection of the source either 
in the MECS
or in the LECS. We derived a 3-$\sigma$ upper limit of
9.4 $\times$ 10$^{-4}$ counts s$^{-1}$ (MECS, 2--10~keV, 42.5 ks).

Based on the angular consistency with the GRB location and the
temporal  behaviour, similar to that of previous GRB afterglows, 
we identify 1SAX J1219.5-0350 as the X-ray afterglow of GRB 990704.

Although the event occurred at high Galactic latitude
and the WFC and NFI error boxes were promptly
searched, no optical or radio afterglow was detected
(\cite{castro99,maury99b,diercks99,vrba99b}).
In the optical R-band the tightest upper limit to a variable object
is at R=22.5 within less than 5 hours after the burst
(\cite{jensen99}) and R$\sim$23.5 (\cite{rol99}) 
after 30.5 hours.

\subsection{Data Analysis}

\begin{figure}
\includegraphics[width=9.5cm]{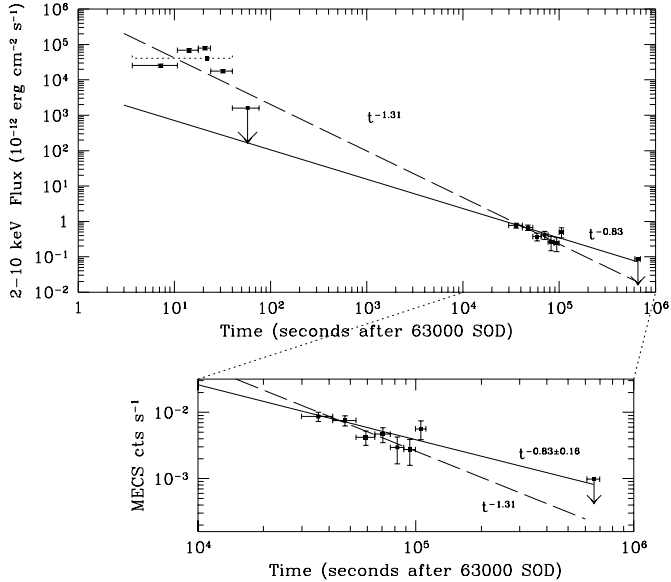}
\vspace{-1.5cm}
\caption{
Decay of the X-ray afterglow of GRB 990704, as energy flux (upper panel)
and photons flux (lower panel).
}
\label{nfi_decay}
\end{figure}

The MECS 2-10~keV light curve of the first observation was accumulated 
with time bins of 11640~s, using an extraction radius of 2' around the 
source centroid. 
The background in the same bandpass was extracted at two different 
positions in the same field, with extraction radii of 2' as well. 
The background is in both cases consistent with a constant value of 
1.8 $\times$ 10$^{-3}$ counts s$^{-1}$
and an analysis over the field shows no systematic trend. 
To derive the MECS background-subtracted light curve in 
Figure~\ref{nfi_decay}, we used the average of the two background 
regions at each time bin. A constant value for the background yelded
consistent results.   
In Figure~\ref{nfi_decay} (bottom panel) we show the temporal 
behaviour of the X-ray flux (2--10~keV) from
1SAX J1219.5-0350 in units of MECS counts s$^{-1}$.

The photon spectrum from the LECS and MECS instruments has been fit
with a simple power law, $I(E)=C E^{-\Gamma}$ 
photons s$^{-1}$ cm$^{-2}$ keV$^{-1}$. The LECS data were
used in the energy range between 0.16 and 4.0 keV, where the response
matrix is most accurate, while the MECS data were used
in 1.6-10.0~keV. The result of the fit is 
$\Gamma$=(1.69$^{+0.60}_{-0.34}$) (reduced $\chi^{2}$=1.83, 17 d.o.f.),
giving an average flux in 2--10~keV of 
(5.0$\pm$0.4) $\times$ 10$^{-13}$ erg cm$^{-2}$ s$^{-1}$
and (2.96$\pm$0.57) $\times$ 10$^{-13}$ erg cm$^{-2}$ s$^{-1}$ 
in 0.16--2.0~keV.
The absorption at low energies is not very well constrained but
appears consistent with that expected 
from absorption by cold interstellar gas along the line of sight
as interpolated from the HI-maps by Dickey \& Lockman (1990)
which results in $N_{\rm H}$=3 $\times$ 10$^{20}$ cm$^{-2}$.
Our spectral results are in agreement with those obtained
by Yonetoku et al. (2000) using ASCA data on the same afterglow.

We also performed the spectral analysis of the LECS and MECS data 
dividing the first observation in two portions of similar duration, 
in order to check 
for spectral evolution. We used an absorbed simple power
law model to fit the data and we got satisfactory results in both
parts. In the first spectrum the best fit value for the power law
photon index is $\Gamma$=(1.78$\pm$0.45) (reduced $\chi^{2}$=1.02, 8 degrees
of freedom). For the second part of the observation we obtained
$\Gamma$=(1.48$^{+0.75}_{-0.67}$) (reduced $\chi^{2}$=1.04, 5 d.o.f.).
Uncertainties on $\Gamma$ are given at 90\% confidence for one parameter.
In both cases the hydrogen absorption column is consistent with the
average Galactic value.

The average MECS spectrum was used to convert the afterglow light
curve from counts s$^{-1}$ to erg cm$^{-2}$ s$^{-1}$. The resulting 
light curve is shown in the top panel of Figure~\ref{nfi_decay}, 
together with the 2-10~keV WFC flux during the GRB prompt event 
averaged over the whole observation (dotted horizontal bar)
and time-resolved in the four
intervals as indicated in Fig.~\ref{lc_wfc_grbm},
and the upper limit obtained from WFC data during the 
$\sim$36~s after the event.
The light curve during the first NFI observation is well described
by a power-law decay $I(t) \propto t ^{-\beta}$, with 
$\beta$ = 0.83$\pm$0.16. However, including in the fit the average
X-ray flux measured by the WFC during the GRB yields $\beta$=1.31.
Both indices are consistent with the NFI upper limit 
during the second observation, but only the flatter one is
consistent with the WFC upper limit
soon after the event. On the other hand, 
the back extrapolation of the law $I(t) \propto t ^{-0.83}$ 
largely underestimates the prompt X-ray emission.

An ASCA observation of the same X-ray afterglow continued 12~ks 
beyond the BeppoSAX one (\cite{murakami99}). 
An early suggestion of a flare in the X-ray flux at the end
of the ASCA observation (\cite{yoshida99a}) was not confirmed
by a later analysis (\cite{murakami00,yonetoku00}).
We searched the last portion of our data for the beginning of such 
a flare and found a marginal indication of flux rise during the 
last two orbits of our observation.
Flaring of the X-ray afterglow has been observed in 
two events in the past: GRB 970508 (\cite{piro99}) and GRB 970828
(\cite{yoshida99b}). In both cases there were hints 
of a redshifted Fe-K emission line in the X-ray spectrum, at the time
of the flare. We searched for spectral features in our spectrum,
both for the whole observation and for its last part.
No evidence is found for an emission line.
Yonetoku et al. (2000) have used the ASCA data on GRB 990704
to set a tight upper limit (equivalent width smaller than $\sim$250 eV) 
to possible Fe-K emission line. 

\section{Discussion}

Several characteristics of GRB 990704 appear noteworthy.
Regarding the GRB prompt emission:

1) The first portion of the event appears dominated by the 
X-ray emission (see Figure 1). 
Only the main peak is distinctively detected 
at energies above 100 keV.

2) The X-ray content of the prompt emission is extremely high, compared to 
the other BeppoSAX GRBs. 
Figures~\ref{ratios_p} and \ref{ratios_s} show how distinct GRB 990704 is 
with respect to the distribution of the other events within 
the BeppoSAX GRB sample mentioned in Sect.~2.  
In fact, the second largest 
value found in our sample for P$_{2-26~keV}$/P$_{40-700~keV}$ 
is 0.45 for GRB981226 (whereas 990704 shows a value of 1.56). 
In terms of fluences,
we find S$_{2-26~keV}$/S$_{40-700~keV}$ = 2.84 for 990704, 2.40 for 
981226 and 0.95 for the largest, 980326, in the remaining sample.
That is, for GRB 990704
{\it the energy recorded in 2--26~keV is almost 3 times that 
recorded in 40--700~keV. } From this point of view, GRB 990704 
hardly meets the classical GRB defining criteria.
Still, it showed an X-ray afterglow as do `classical' GRBs.

3) The main peak observed in gamma-rays (i.e., the third one in X-rays)
appears to be consistent with the typical energy dependence of the pulse width 
($\Delta$t $\propto$ E$^{-0.45}$) observed in many individual GRB
pulses (\cite{fenimore95}). This dependence is consistent with a 
scenario in which the
prompt emission is due to synchrotron radiation in a shocked medium
(e.g., \cite{tavani96}). From this point of view GRB 990704 appears to be a
`classical' GRB.

The X-ray afterglow of GRB 990704 is peculiar.
The BeppoSAX measurement shows a very slow decay, with index $\sim$0.8.
If we consider a sample
of thirteen BeppoSAX X-ray afterglows (Stratta et al., in preparation) 
from 1997 to May 1999, we find that the average value of the power law 
decay slopes is 1.36 and the standard deviation is only 0.16. 
However, a word of caution: a power law decay with 
a slope of only 0.8, if extrapolated to an infinite time would cause
the total emitted energy to diverge. Therefore, we have to assume 
that {\it after} the end of the BeppoSAX observation the decay slope 
changed to a larger value.
In this respect, recall the claim (but not confirmed by a later analysis)
of an ASCA detection of rebursting of the X-ray flux from this source in 
the few hours following the end of the BeppoSAX observation.
If such a rebursting occurred, 
then the slow decay measured by BeppoSAX could
be an artifact of an incomplete sampling of the bursting behaviour 
of the X-ray flux.
In this case the true overall decay would be indeed faster, 
possibly the $\sim$1.3 slope we obtain taking into account 
the prompt X-ray emission.

The softness of the prompt emission from GRB 990704 is definitely 
peculiar. As shown in Figure~\ref{ratios_s}, a similarly large 
X-to-gamma energy ratio was only observed once 
in the BeppoSAX GRB sample, in the case of GRB 981226.
It too was peculiar in its X--ray afterglow, showing an initial 
short rise, or maybe a plateau, never seen in any other X-ray afterglow
(\cite{frontera00}).
In the GINGA sample of 22 GRBs (\cite{strohmayer98}), only one event,
GRB 900901,
had a X-to-$\gamma$ fluence ratio comparable to GRB 990704. In that case
the ratio was 1.23, although 
the gamma-ray fluence was measured in 50-300 keV, and this most likely
slightly biases upward the value of the ratio when compared to our sample.

In a few other BeppoSAX 
GRBs the ratio of X- to gamma-ray fluence was relatively
high (up to 0.4) with respect to the less than $\sim$10\% 
usually observed in GRBs. For example,
see Table 3 in \cite{frontera00b} and Figure~\ref{ratios_s} for the 
BeppoSAX events, and Table 1 in Strohmayer et al. 1998 for GINGA, 
where the average value of S$_{2-10~keV}$/S$_{50-300~keV}$ goes down to 
$\sim$12\% if one excludes just the largest 3 values.
It is intriguing to note that the events in the BeppoSAX sample with a 
high X-ray content in their prompt emission include all five 
GRBs proposed to be associated with supernovae, on the basis of the
properties of their optical afterglow.
In fact, they showed the following values for 
S$_{2-10~keV}$/S$_{40-700~keV}$ (\cite{frontera00b,frontera01}): 
980425: 28\%; 980326: 38\%; 970228: 20\%; 990712: 40\%; and 970508: 20\%.
Taken individually, these are not extraordinary values but it is suggestive
that {\it all} of them have values larger than the average. 
Actually, as Frontera et al. (2000a) suggested, the explosion of a 
supernova prior to the GRB may enrich the circumburst 
medium with baryon-rich matter. This would provide a natural explanation
for a lower Lorentz factor of the expanding fireball following the
GRB explosion, leading to the observed lower peak energy of the emitted 
radiation and consequently the X-ray richness in this type of GRBs.
We note, however, that in some of the cases mentioned above 
(e.g., 980425 and
980326) the assumption is that the supernova exploded right at the 
time of the GRB, so that this scenario may not apply in a straightforward
way.

Finally, we note that GRB 990704 is definitely the BeppoSAX GRB with 
characteristics that are
the closest to the recently suggested class of FXTs.
As mentioned in the introductory section, these events are not yet
firmly characterized. Based on the BeppoSAX data (\cite{heise01}), 
so far we know that
they are almost indistinguishable from X-ray counterparts of classical 
GRBs (in terms of duration, hardness, flux),
but they do not show gamma-ray emission detectable by the GRBM.
Quantitatively, the X-to-gamma peak flux ratios for FXTs go
from few to a hundred, while the ratio of fluences goes from 0.1 to 10
(\cite{heise01}).
Therefore, GRB 990704 appears as the classical GRB
(because it has a GRBM-detected emission and an X-ray afterglow)
within the BeppoSAX sample that most closely approaches the properties
of the FXTs.
Should the X-ray content be the signature of a supernova, then 
supernova explosions would be good candidates for the origin of FXTs.

More observations will be needed to give
(or not) observational support to our conjectures but the current data 
are intriguing, and theoreticians are already at work 
on similar ideas, e.g., Woosley (2001).

\section{Acknowledgements}
The authors would like to remember the key role 
of Daniele Dal Fiume in the BeppoSAX project, and the
GRB project in particular. They are grateful to
Chris Butler, former BeppoSAX Mission Director, and to the
Teams at the BeppoSAX Science Operation Center, Operation Control 
Center and Science Data Center for their enthusiastic contribution to 
the BeppoSAX GRB program.
The authors warmly thank Hale Bradt for suggestions and for a careful 
reading of the manuscript.
They also thank the anonymous referee for comments that improved the paper.
BeppoSAX is a program of the Italian Space Agency (ASI) with participation
of the Netherlands Agency for Aerospace Programs (NIVR).

\end{document}